\documentclass{eptcs}

\usepackage{url}
\usepackage{wrapfig,caption,subcaption}
\usepackage{amsmath,amssymb,amsthm}
\usepackage{xspace,xcolor}
\usepackage{paralist}

\usepackage{tikz}
\usetikzlibrary{arrows}
\usetikzlibrary{shapes}
\usepackage{tabularx} 

\usepackage{basics}
\usepackage[nobook]{theorems}
\usepackage[lfsyntax]{twelf-math}
\usepackage{local}

\usepackage{MnSymbol}

\begin{document}

\title{A Logic-Independent IDE}
\def\titlerunning{A Logic-Independent IDE}

\author{Florian Rabe\institute{Computer Science, Jacobs University, Bremen, Germany \\ \email{f.rabe@jacobs-university.de}}}
\def\authorrunning{Florian Rabe}
\maketitle

\begin{abstract}
The author's MMT system provides a framework for defining and implementing logical systems.
By combining MMT with the jEdit text editor, we obtain a logic-independent IDE.
The IDE functionality includes advanced features such as context-sensitive auto-completion, search, and change management.
\end{abstract}

\section{Introduction and Related Work}\label{sec:intro}
  \paragraph{Motivation}
Deduction systems such as type checkers, proof assistants, or theorem provers have initially focused on soundness and efficiency.
Consequently, the most natural design choice for the user interface (UI) has often been a kernel that implements the logic and the theorem prover and with which users interact through a read-eval-print loop.
But over time formalization projects have reached larger scales that call for more sophisticated UIs, and it has proved non-trivial to add these to existing kernels.

Users more and more expect interactive, flexible UIs, e.g., like the ones they know from software engineering.
They ask for features such as highlighting of errors, fast rechecking after editing any part of the project, or a powerful search function over the library.
For example, in personal communications with the author, members of the Feit-Thompson project in Coq \cite{oddorder} pointed out the UI as a key bottleneck (rather than the type theory or the theorem prover); and Georges Gonthier expressed his wish for a better UI for searching.
Similarly, members of the L4.verified project in Isabelle \cite{l4verified_mkm} urged for better change management support because their work was often delayed by long rechecking cycles.
And for casual users from outside the theorem proving community, the UI can be more important than the underlying logic.

However, developers' resources tend to be stretched already by developing (and maintaining) the kernel at all.
Therefore, it is no coincidence that none but the most mature systems like Coq \cite{coq}, Isabelle \cite{isabelle}, or ACL2 \cite{acl2sedan} are equipped with graphical UIs.
Arguably the most powerful UI is Makarius Wenzel's Isabelle/\jedit \cite{isabelle_jedit}, and its history is telling: It required a multi-year effort by one of the lead developers, throughout which he had to reimplement parts of the kernel in order to expose the needed kernel functionality to the UI component.
\medskip

In this situation, it is important to investigate whether powerful UIs can be realized independently of the kernel.
This is well-known to be the case in software engineering, where IDE frameworks like Eclipse \cite{eclipse} are parametric in the programming language, which is supplied as a plugin.
This separation of concerns between kernel and UI development would be even more valuable in theorem proving, where developer communities are comparatively small.

\paragraph{Related Work}
The idea of generic UIs for theorem proving has so far mainly been applied to \textbf{low-level} user interfaces.
These typically interact with an abstract kernel through a read-eval-print interface.
This has the advantage that almost any kernel can be plugged into the UI at relatively little cost.
Examples are Proof General \cite{proof_general}, which uses a text editor, and Proof Web \cite{proofweb}, which uses a web browser to host the UI.
Similarly, \cite{largeformalwikis} provides a wiki-like frontend for Coq and Mizar.

It is often possible to enrich this abstraction layer with configuration and query commands.
But the low-level approach is inherently limited by the fact that the UI is not aware of the abstract syntax tree (AST) that is computed and refined by the kernel.

We speak of \textbf{high-level} UIs if the UI is aware of the AST and nodes carry cross-references to the corresponding regions in the source.
This enables IDE-style UI features such as AST display, error highlighting, tool tips, hyperlinks, and auto-completion.
The AST may also include the results of, e.g., disambiguation, type reconstruction, and theorem proving.

Moreover, if the kernel exposes functions for rechecking fragments (and their dependencies), high-level UIs can support change management very well.
In low-level UIs, change management is usually limited to rerunning the kernel starting from a previous check point, either declared manually, precomputed as a heap image, or generated automatically when entering a new file.
All of these methods behave poorly when making a change in a file that is read early on, e.g., by tweaking a definition or by adding a lemma about a definition.
For example, the \cite{l4verified_mkm} developers routinely added theorems to the current Isabelle file even if they conceptually belonged into earlier files because moving them to an earlier file incurred high rechecking penalties.

The disadvantage of high-level UIs is that it becomes much harder to plug in an existing kernel (especially if they are implemented in a different programming language than the UI).
In fact, it is not always obvious whether it would be easier to develop a high-level UI for a specific existing kernel or to connect a generic high-level one.

A compromise can be to develop UIs for logical frameworks like Isabelle \cite{isabelle}, which can be inherited by all object logics.
In that sense, Isabelle/\jedit \cite{isabelle_jedit} is a generic high-level UI, and the only one we are aware of besides ours.
We will relate the two in Sect.~\ref{sec:conc}.
A similar argument applies to UIs for very rich systems like Coq \cite{coq}, whose logic can be customized by disabling language features or adding axioms.
The Agda kernel \cite{agda} achieves advanced UI features by dynamically generating code that is evaluated by the Emacs-based UI.

\paragraph{Our Approach}
We hold that in the long run, a decoupling of kernel and UI can prove beneficial if only because there are a lot more potential UI than kernel developers.
Somewhat provocatively, we observe that in monolithic systems, the few people who understand the kernel well enough to improve the UI mostly do not have the time and incentive to do so.

Therefore, we introduce \jmmt, a logic-independent high-level user interface.
Our key idea is to design a rich abstraction layer between kernel and UI.
This permits the logic-independent development of UI features, which results in a valuable separation of efforts:
\begin{compactitem}
	\item Logic developers can focus on designing the logic and implementing the kernel.
	  This frees valuable resources to improve the theorem prover.
	\item UI developers can focus on developing UI features without having to understand any logic or prover.
	  They benefit twice because their work becomes easier and more reusable.
	\item Finally, this benefits developers of auxiliary components like proof hint generators (e.g., \cite{hollight_hints}) or search engines (e.g., \cite{mathwebsearch}).
	  They are enabled to make their services logic-independent and directly integrate them with the UI.
\end{compactitem}

We base our abstraction layer on the \mmt language \cite{RK:mmt:10}, a logic-indepen\-dent representation language that admits natural embeddings of virtually all declarative languages.
We model a kernel as a set of components for parsing and validating the structure of a document as well as the terms within it.
This lets us develop advanced UI features such as change management and search in addition to the ones we mentioned above.

On the implementation side, our IDE makes use of the \mmt API \cite{rabe:mmtabs:13}, which already offers a high-level machine interface to \mmt content, as well as the \jedit text editor, which offers a rich plugin framework for both UI and language-specific extensions.
To connect them, we develop a plugin for \jedit based on \mmt, which implements our abstraction layer and the UI functionality.

To evaluate and exemplify this design, we have developed a kernel for the logical framework LF \cite{lf} that instantiates our abstraction layer.
This yields a powerful IDE for LF, whose UI features surpass existing implementations of comparable languages.

Moreover, our LF kernel is designed to be highly extensible.
Thus, it is possible to develop additional kernels quickly by plugging in, e.g., different operators, notations, or typing rules.


\paragraph{Overview}
Sect.~\ref{sec:prelim} summarizes the preliminaries we need about \jedit and \mmt.
In Sect.~\ref{sec:features}, we describe the UI functionality, which amounts to the user perspective of \jmmt.
In Sect.~\ref{sec:design}, we describe our abstraction layer, which amounts to the kernel developer perspective of \jmmt.

Source code, binaries, documentation, and additional screenshots and videos are available from the \mmt homepage\footnote{\url{https://svn.kwarc.info/repos/MMT/doc/html/index.html}}.
Early parts of this work were carried out with Mihnea Iancu and presented as work-in-progress \cite{IR:ui:12}.

\section{Preliminaries}\label{sec:prelim}
  \subsection{The \jedit Text Editor}
    jEdit is a widely-used Java-based text editor \cite{jedit}.
It is particularly interesting as a UI for formal systems due to its strong plugin infrastructure that can be used to provide IDE-like functionality.
Thus, it provides a lightweight alternative to complex IDE frameworks like Eclipse \cite{eclipse}.

Existing plugins already provide abstract interfaces for among others outline view, error highlighting, auto-completion, hyper-linking, and shell integration that can be implemented by format-specific plugins.
Thus, relatively little glue code is necessary to connect the UI components and the language-specific data model.
  \subsection{The MMT Representation Language}
    \mmt \cite{RK:mmt:10} is a prototypical declarative language.
It systematically avoids committing to an individual logic or type theory and focuses on capturing their joint structural properties.
Specific language features such as function types or equality are defined in separate modules, which serve as building blocks to compose languages.

\begin{wrapfigure}{r}{6.6cm}
	\vspace{-3em}
	\begin{center}
		\begin{tabular}{|l@{\hspace{.3cm}}l@{\hspace{.3cm}}l@{\hspace{.3cm}}l|}
			\hline 
			Theory                 & $\Sigma$    & $\bbc$ & $\cdot \alt \Sigma,\; c\opt{:E}\opt{=E}\opt{\#N}$ \\
			Context                & $\Gamma$    & $\bbc$ & $\cdot \alt \Gamma,\; x:E$ \\
			Term                   & $E$         & $\bbc$ & $c \alt x \alt \cons{c}{\Gamma}{\rep{E}}$ \\
			Notation               & $N$         & $\bbc$ & $\rep{(\vmk{n}\alt \amk{n}\alt \mathtt{string})}$\\ 
			\hline
		\end{tabular}
		\caption{\mmt Grammar}\label{fig:mmt-grammar}
	\end{center}
\vspace{-2em}
\end{wrapfigure}

Here we introduce only a small fragment of \mmt that is sufficient to exemplify our IDE.
The grammar is given in Fig.~\ref{fig:mmt-grammar}.
A \textbf{theory} $\Sigma$ is a list of \textbf{constant} declarations.
A \textbf{constant} declaration is of the form $c[:A][=t][\#N]$ where $c$ is an identifier, $A$ is its type, $t$ its definiens, and $N$ its notation, all of which are optional.
A \textbf{context} $\Gamma$ is very similar to a theory and declares typed variables.

Type and definiens are \textbf{terms}, which are formed from the constants, variables, and complex terms.
We use a general form $\cons{c}{\Gamma}{E_1,\ldots,E_n}$ for complex terms, which subsumes binding and application: $c$ is called the \textbf{head} of the term, the (possibly empty) context $\Gamma$ declares the \textbf{bound variables}, and the $E_i$ are the \textbf{arguments}.
For example, in a $\lambda$-term $\cons{\lflambda}{x:A}{t}$, $\lambda$ is the head, $x:A$ the bound variable context, and $t$ the single argument.

\begin{wrapfigure}{r}{4cm}
   \vspace{-1em}
	$\begin{array}{l@{\tb\#\tb}l}
	\type     & \type \\
	\kind     & \kind \\
	\lfPi     & \{\,\vmk{1}\,\}\,\amk{2},\\
	\lflambda & [\,\vmk{1}\,]\,\amk{2},\\
	\lfapply  & \amk{1}\,\;\,\amk{2}, \\
	\lfarrow  & \amk{1}\to\amk{2}\\
	\end{array}$
	\caption{LF in \mmt}\label{fig:lf}
\vspace{-.5em}
\end{wrapfigure}

The notation of $c$ is used for the concrete representation of complex terms with head $c$: $\vmk{n}$ refers to the $n$-th bound variable, $\amk{n}$ to the $n$-th argument (counting starts from the last variable), and they may be interspersed with arbitrary delimiters. 
For example, if we declare the constant $\lflambda$ with the notation $\lambda\,\vmk{1}\,.\,\amk{2}$, then $\lambda x:A.t$ becomes the concrete representation of $\cons{\lflambda}{x:A}{t}$.

The following theories will serve as running examples throughout the paper:

\begin{example}[LF as an \mmt Theory]\label{ex:lf}
	To obtain the logical framework LF, we use the theory shown in Fig.~\ref{fig:lf}.
These constants do not have types.
Instead, they constitute primitive concepts, which can be used to form terms, which can then occur as the types of other constants.
However, their binding-arity and argument-arity can be inferred from their notations, e.g., $\lfPi$ binds one variable and then takes one argument.
\end{example}

Once LF is defined, we can use it as a logical framework to define other logics:

\begin{example}[A Logic in LF]\label{ex:pl}
Using the theory from Ex.~\ref{ex:lf}, we can define a theory for a simple logic by declaring (among others) the following constants
\[\begin{array}{l@{\;:\;}l@{\;\#\;}l}
  \prop     & \type & \prop, \\
  \ded      & \prop\to\type & \prop, \\
  \lfop{imp} & \prop\to\prop\to\prop & \amk{1}\Arr\amk{2,}\\
  \lfop{and} & \prop\to\prop\to\prop & \amk{1}\wedge\amk{2},\\
  \lfop{andI}& \{A\}\{B\}\,\ded\, A\to\ded\,B\to\ded\,(A\wedge B) & \lfop{andI}\,\amk{3}\,\amk{4},\\
  \lfop{impI}& \{A\}\{B\}\,(\ded\,A\to\ded\,B)\to\ded\,(A\Arr B) & \lfop{impI}\,\amk{3}\\
  \lfop{example} & \multicolumn{2}{l}{\{A\}\ded\,(A\Arr (A\wedge A)) \;=\; [A] \lfop{impI}\,[p]\,\lfop{andI}\,p\,p}
\end{array}\]

This theory uses the Curry-Howard representation of judgments as types: The provability judgment of $A:\prop$ is represented as the term $\ded\,A$.
The constant $\mathtt{example}$ declares and proves a (trivial) theorem using the natural deduction rules for conjunction and implication introduction.
We already omit inferable variable types and declare arguments to be implicit by not mentioning them in the notation (e.g, the first $2$ arguments of $\mathtt{impI}$ are implicit). 
\end{example}

\mmt abstracts from the typing relation between terms.
It fixes only the judgments about terms as given in Fig.~\ref{fig:judge}.
These judgments can be defined arbitrarily by specific languages represented in \mmt.

\begin{figure}
	\begin{tabularx}{\textwidth}{|l|X|}
		\hline
		 $\isuniv{T}{A}$    & inhabitability, i.e., $A$ may occur as the type of a constant \\
		 $\oftype{T}{E}{A}$ & typing relation, in particular $E$ may occur as the definiens of a constant with type $A$ \\
		 $\isequal[A]{T}{E}{E'}$ & equality of terms at type $A$\\
		\hline
	\end{tabularx}
 \caption{\mmt Judgments Relative to a Theory with Name $T$}\label{fig:judge}
\end{figure}

The \mmt \textbf{implementation} \cite{rabe:mmtabs:13} provides an API for maintaining \mmt content.
It is written in Scala and thus fully interoperable with the Java codebase of \jedit.
It is not an application with its own user interface.
Instead, it focuses on providing services that can be used in \mmt-based applications (such as our IDE).

The \mmt implementation is highly extensible and relegates many features to plugin interfaces.
In particular, to instantiate \mmt with specific languages, one supplies a plugin that implements the judgments from Fig.~\ref{fig:judge}.

Note that via Curry-Howard, \textbf{axioms}, \textbf{theorems}, \textbf{tactics}, \textbf{inference rules}, etc. can be represented as \mmt constants.
In particular, the type of a theorem represents the asserted formula and its definiens the proof (as in the constant $\mathtt{example}$ of Ex.~\ref{ex:pl}).
In that case, the typing relation represents the correctness of proofs.

This is flexible enough to represent the tactic-based proofs used in interactive theorem provers: For example, we can declare an untyped constant $\mathtt{auto}$ and use it as a proof.
In that case, the implementation of the typing judgment would include the theorem proving necessary to determine whether $\mathtt{auto}$ can be accepted as a proof.

\section{Generic User Interface Functionality}\label{sec:features}
  \subsection{Basic Features}\label{sec:basic}

We use the phrase ``basic features'' to group together features that are relatively easy to realize on top of the \jmmt abstraction layer.
Most of these features make use of existing \jedit UI functionality, specifically the \jedit plugins SideKick, ErrorList, and Hyperlinks.
This is not, however, meant to imply that these features are inherently easy to realize.
Instead, this easiness positively evaluates our design.

\begin{figure}
	\begin{center}
		\includegraphics[scale=0.385]{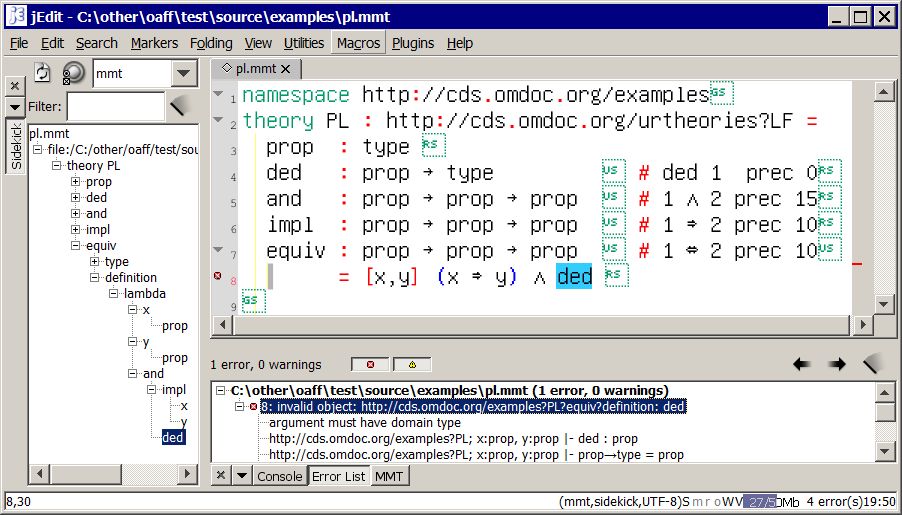}
	\end{center}
	\vspace{-1.5em}
	\caption{\jmmt Instantiated with Our LF Kernel}\label{fig:jedit}
\end{figure}

\paragraph{Abstract Syntax Display}
A dockable subwindow displays the abstract syntax tree (AST) of the current buffer.
This is shown on the left of Fig.~\ref{fig:jedit}, where we define the theory from Ex.~\ref{ex:pl} in LF.
Here, the green parts are the keywords and separators of our structure parser (see Sect.~\ref{sec:sp}).

Users can navigate from a buffer position to the respective node in the AST and vice versa, as shown in the jointly selected occurrence of $\mathtt{ded}$.
This is achieved by using \textbf{source references}: Every node in the AST points to the region it covers in the buffer.
An example for the strength of source references in the AST is that it took the author $< 15$ minutes to code the following immensely useful feature: Double-clicking on an operator selects the smallest sub-term containing it.

The AST also includes information that is not present in the source but inferred by the kernel.
This can be seen in the types $\mathtt{prop}$ of $\mathtt{x}$ and $\mathtt{y}$, which occur in the AST but not in the source.
Similarly, all inferred implicit arguments occur in the AST.

\paragraph{Error Highlighting}
All errors returned by the kernel are displayed in a dockable window and via line markers.
This is shown at the bottom of Fig.~\ref{fig:jedit}.
There the kernel detected an example typing error in the definition of the equivalence connective: $\ded$ is used instead of $y\Arr x$.
Clicking on the error message resulted in the selection of this ill-typed sub-term.

The error message also includes the kernel's log messages from the branch of the derivation that led to the error.
In this case, the typing judgment $\ded:\prop$ failed.

Note that the AST display is robust against errors.
Firstly, parsing of the ill-typed term succeeded so that the AST can be displayed.
Secondly, even though type-checking of the term failed, the type-checker returned a partially checked term.
We can see that in the inferred types of the bound variables $\mathtt{x}$ and $\mathtt{y}$, which are displayed correctly in the AST even though they occur in an overall ill-typed term.
This is very important because UI support is needed especially when fixing errors.

\paragraph{Tooltips and Hyperlinking}
For all occurrences of variables and constants, \jmmt displays their types as tooltips.
This includes the inferred types of bound variables and the inferred implicit arguments of constants.

Moreover, users can navigate from every occurrence of a constant to its declaration.
Because the \mmt project manager (see Sect.~\ref{sec:projects}) serializes the AST of every source file to disk, this includes declarations in files or projects that are not currently open in the IDE.

\subsection{Auto-Completion and Proof Hints}\label{sec:hints}

\jmmt provides a basic auto-completion feature (realized based on SideKick).
Using the AST, it determines the current theory and context at the cursor position and suggests identifiers that are in scope.

	\begin{wrapfigure}{r}{6.5cm}
		\vspace{-2em}
		\begin{center}
			\includegraphics[scale=0.6]{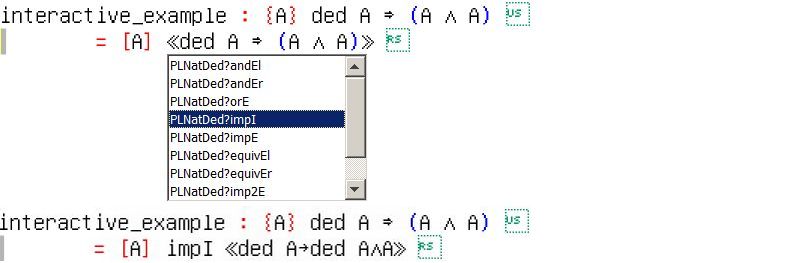}
		\end{center}
		\vspace{-1.5em}
		\caption{Proof Hints}\label{fig:complete}
		\vspace{-1em}
	\end{wrapfigure}
	
Moreover, we realize a general hinting feature as follows.
We declare a additional \mmt constant and implement a typing rule for it such that the term $\llangle E\rrangle$ is inferred to have type $E$.
Both users and plugins can use such terms to represent a missing term of type $E$ (e.g., an open subgoal).
In that case, \jmmt is aware of the expected type at the cursor position and can provide better auto-completion support.

\begin{example}\label{ex:complete}
	In our kernel for LF, we implement a completion rule:
	If \jmmt asks for completions for the open goal $\llangle E\rrangle$, this rule finds all constants or variables $c$ of type $\{x_1\}\ldots \{x_m\}T_1\to\ldots\to T_n\to T$ such that $T$ unifies with $E$ via substitution $\sigma$.
	If a completion is chosen by the user, the rule inserts $c\;\sigma(x_1)\;\ldots\;\sigma(x_m)\;\llangle \sigma(T_1)\rrangle\;\ldots \llangle \sigma(T_n)\rrangle$.
\end{example}

In the upper part of Fig.~\ref{fig:complete}, the user is about to choose implication introduction towards proving the example theorem from Ex.~\ref{ex:pl}.
Our rule returns $\mathtt{impI}\,A\,(A\wedge A)\,\llangle \ded\,A\to\ded\,A\wedge A\rrangle$.
The bottom part shows the situation afterwards (where the implicit arguments have been removed when inserting the hint into the source).
  
Using a second rule to introduce $\lambda$-terms, the whole example proof can be completed using repeated auto-completion.

More generally, any tactic, decision procedure, or proof hinting algorithm can be realized as an auto-completion in this way.

\subsection{Interactive Type Inference}\label{sec:infer}

\begin{figure}[t]
  \begin{subfigure}[t]{0.5\textwidth}
   \begin{center}
		\includegraphics[scale=0.5]{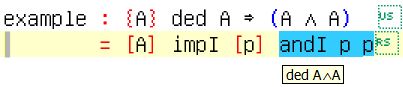}
		\caption{Type Inference}\label{fig:infer}
   \end{center}
  \end{subfigure}
  \begin{subfigure}[t]{0.5\textwidth}
   \begin{center}
		\includegraphics[scale=0.5]{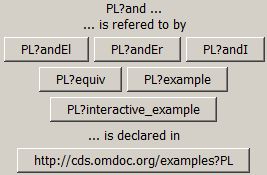}\\[.2cm]
		\caption{Navigation}\label{fig:navigate}
   \end{center}
  \end{subfigure}
  \bigskip
  
  \begin{subfigure}[t]{0.5\textwidth}
   \begin{center}
		\includegraphics[scale=0.5]{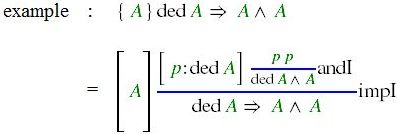} \tb\tb\tb
		\caption{Tree View}\label{fig:web-tree}
   \end{center}
  \end{subfigure}
  \begin{subfigure}[t]{0.5\textwidth}
   \begin{center}
		\includegraphics[scale=0.5]{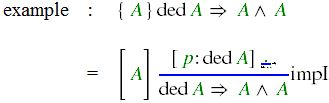}
		\caption{Folding}\label{fig:web-fold}
   \end{center}
  \end{subfigure}
\caption{\jmmt Screenshots}\label{fig:screenshots}
\end{figure}

If a subterm is selected, \jmmt queries the kernel for its type and displays it as a tooltip.
This can be seen in Fig.~\ref{fig:infer} where the user selected a sub-term of the proof of the theorem from Ex.~\ref{ex:pl}.

\subsection{Project Management and HTML View}\label{sec:projects}

\mmt provides a notion of projects \cite{HIJKR:dimensions:11} modeled after projects in software engineering.
Among other things, this includes a build tool that calls the kernel on all files in a project and caches the resulting ASTs to disk using the \omdoc XML format \cite{omdoc}.
Therefore, \jmmt can access all dependencies of the current source file without them being processed during the current session.

The \mmt build tool can also produce a documentation website for the project, which includes a hierarchical and graph-based project browser.
For example, Fig.~\ref{fig:web-tree} shows the theorem from Ex.~\ref{ex:pl} as rendered in a web browser.

Moreover, \jmmt automatically starts a web server, through which the project pages can be browsed interactively.
Interactive features include selecting or folding sub-terms (as in Fig.~\ref{fig:web-fold}) or showing/hiding any inferred part of the term.
The user can also control \jmmt via the web browser; in particular, by clicking on a declaration in the web browser, the user can open it in \jedit.

\subsection{Relational Navigation}

The \mmt build tool also produces a relational index that represents an ontology-style abstraction of an \mmt document.
For example, atomic relations in the index are the import-relation between theories and the occurs-in relation between parsing units.
\mmt includes a query language over this ontology \cite{rabe:querying:12}, which can, e.g., take the transitive closure of a relation or its restriction to theorems.

A particularly useful application in \jmmt is relational navigation: It displays one navigation button for every declaration that is related to the current one.
Clicking a button navigates to that declaration.
This permits very fast navigation across a project.
Fig.~\ref{fig:navigate} shows for example all constants that refer to $\mathtt{and}$.

\subsection{Search}\label{sec:search}

\begin{wrapfigure}{r}{8cm}
	\vspace{-4em}
	\begin{center}
		\includegraphics[scale=0.4]{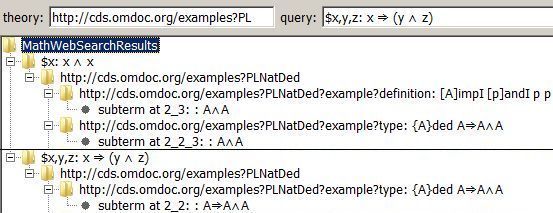}
	\end{center}
	\vspace{-1.5em}
	\caption{Display of Search Results}\label{fig:search}
	\vspace{-2em}
\end{wrapfigure}

The \mmt build tool also creates an index of all terms in a project.
This index can be read by the MathWebSearch tool \cite{mathwebsearch}, which uses substitution tree indexing to make the set of terms available for fast searching.
If an instance of MathWebSearch is running, it can be registered with \jmmt, which then provides a search interface for the project.

In Fig.~\ref{fig:search}, \jmmt displays the results of two search queries, where only results from the running example are shown.
A search query is of the form $\$x_1,\ldots,x_n:E(x_1,\ldots,x_n)$ and returns all terms that unify with $E(x_1,\ldots,x_n)$.
The first query asked for all terms of the form $x\wedge x$ for arbitrary $x$, and the second for all terms of the form $x\Arr (y\wedge z)$.

Note that the results include hits in the inferred parts of the term: For example, the term $A\wedge A$ is found in the definiens $[A]\,\mathtt{impI}\,[p]\,\mathtt{andI}\,p\,p$ because it occurs in the implicit arguments of $\mathtt{impI}$.

Clicking on a result opens the containing file in \jedit and selects the respective sub-term.

\subsection{Change Management}\label{sec:moc}

The validation of terms is typically the most expensive part of checking a file because it can involve theorem proving or computation.
Therefore, it is desirable to recheck terms only when necessary.

\begin{wrapfigure}{r}{5.9cm}
	\vspace{-1.5em}
	\begin{tikzpicture}[scale=0.8]
	\node (s)  at (0,3.5) {$t$};
	\node (st) at (0,3) {string};
	\node (sp) at (0,2) {parsed};
	\node (sv) at (0,1) {validated};
	
	\node (t)  at (3,3.5) {$t'$};
	\node (tt) at (3,3) {string};
	\node (tp) at (3,2) {parsed};
	\node (tv) at (3,1) {validated};
	
	\node (dots) at (6,1) {\ldots};
	
	\draw[\arrowtip-] (st) -- (sp);
	\draw[\arrowtip-] (sp) -- (sv);
	\draw[\arrowtip-] (tt) -- (tp);
	\draw[\arrowtip-] (tp) -- (tv);
	
	\draw[\arrowtip-] (sv) -- (tv);  
	\draw[\arrowtip-] (tv) -- (dots);  
	
	\draw[-\arrowtip] (4.5,2.5) --node[above]{depends on} (5.5,2.5);  
	\end{tikzpicture}
	\vspace{-3em}
\end{wrapfigure}

Therefore, \mmt maintains a two-dimensional dependency graph for all terms (i.e., all types/definientia of all constants) in a project as indicated on the right.
In the vertical dimension, every term is maintained in three forms: the string in the source file, the result of parsing it, and the result of validating it.
In the horizontal dimension, \mmt maintains a logical dependency relation between terms.

These horizontal dependencies are known from software engineering, where IDEs recompile a source file only when necessary.
For logics, it is important to use a more fine-granular dependency relation.
For example, \cite{dependencyrelations} extract fine-granular dependencies for Coq and Mizar.
We use terms as the granularity level: If the type or definiens of a constant is used while checking another one, this creates a horizontal dependency.
\medskip

For example, consider two theorems $c:A=p$ and $d:B=q$ such that $d$ uses $c$.
Then type checking the proof of $d$, i.e., $q$, has to look up the formula asserted by $c$, i.e., $A$.
Thus, we obtain a horizontal dependency from $q$ to $A$.
Let us now assume that we make a change in $p$ without changing $A$, e.g., we might fix an error in the proof.

Without change management, \jmmt would have to ask the kernel to recheck the whole buffer and then replace the old AST with the new one returned by the kernel.
This may even require rechecking multiple files (which is what caused the delays in the L4.verified project we cited in Sect.~\ref{sec:intro}).

As we will describe in Sect.~\ref{sec:design}, \jmmt splits the kernel conceptually into several components such that every term can be parsed and validated individually.
This permits \jmmt to patch the AST by calling the kernel only on the necessary terms.
More precisely, a term is reparsed only if its string representation has changed.
If this causes the parsed representation to change, the term is revalidated as well.
For example, this is not the case if we only changed a comment or the formatting of a term.
Finally, if a term has to be revalidated and the resulting validated representation has changed, we also revalidate all terms that horizontally depended on it.

For our two-theorem example, that means that the changes in $p$ do not trigger the expensive revalidation of $q$.
This is crucial to achieve short edit-check cycles when working with multiple theorems at the same time.

\section{An Abstraction Layer Between Kernel and UI}\label{sec:design}
\begin{wrapfigure}{r}{7.5cm}
		\vspace{-2em}
\begin{tikzpicture}[data/.style={rectangle,draw=black,rounded corners}]
  \node[data] (U) at (-2,4) {user interface};
  \node[data] (T) at (-4,2) {\begin{tabular}{c}text \\ representation\end{tabular}};
  \node[data] (M) at (0,2) {\begin{tabular}{c}\mmt\\ representation\end{tabular}};
  \draw (-5.6,1.2) rectangle (1.7,4.3) node[above=0.2cm,left] {\jmmt IDE};
  \node[data] (p) at (-4,-0.5) {\begin{tabular}{c}structure parser \\ \hline term parser\end{tabular}};
  \node[data] (v) at (0,-0.5) {\begin{tabular}{c}structure validator \\ \hline term validator\end{tabular}};
  \draw (-5.6,-1.2) rectangle (1.7,0.4) node[below=1.8cm,left] {kernel};
  \draw[-\arrowtip] (T) --node[right] {parsing} (p);
  \draw[-\arrowtip] (p) -- (M);
  \draw[-\arrowtip] (M) to[out=-135,in=135] (v);
  \draw[-\arrowtip] (v) to[out=45,in=-45] node[left] {validation} (M);
  \draw[-\arrowtip] (U) --node[left] {uses} (T);
  \draw[-\arrowtip] (U) --node[right=-.1cm] {uses} (M);
  \end{tikzpicture}
	\vspace{-2em}
\end{wrapfigure}

An \textbf{overview} of architecture is shown on the right, where the abstraction layer separates \jmmt from a kernel.
Both the concrete text source and its abstract \mmt representation are maintained and used by \jmmt.
\jmmt uses abstract operations for parsing and validation, which are provided by kernels.

\begin{wrapfigure}{r}{7cm}
	\vspace{-2.5em}
	\begin{tabular}{|l||ll|}
		\hline
		$2\times 2$ operations & Parsing & Validation \\
		\hline
		\hline
		Structure & Sect.~\ref{sec:sp} & Sect.~\ref{sec:sv} \\
		Terms     & Sect.~\ref{sec:tp} & Sect.~\ref{sec:tv} \\
		\hline
	\end{tabular}
	\vspace{-1em}
\end{wrapfigure}


More concretely, the abstraction layer consists of the $2\times 2$ operations shown on the right.
Along one dimension, we distinguish the \textbf{structure} and \textbf{term} levels.
The structure level comprises the names of the theory and constant declarations.
This corresponds to the OCaml toplevel in HOL Light \cite{hollight} or to the outer syntax of Isabelle.
The term level comprises the type and definiens of the constant declarations.
This corresponds to the HOL Light formula parser or to the inner syntax of Isabelle.

Along the second dimension, we distinguish \textbf{parsing} and \textbf{validation}.
The former produces an abstract syntax tree that closely resembles the source text;
the latter refines this AST using advanced operations such as type checking, theorem proving, or computation.
Notably, we use \mmt to represent both the parsed and the validated representation, i.e., validation is a transformation from \mmt syntax to \mmt syntax.


A major \textbf{motivation} of this separation is to isolate the validation of terms.
This has three reasons.
Firstly, this is the most expensive phase.
Therefore, as we described in the change management section, isolating this component permits revalidating terms as rarely as possible.
Secondly, almost all errors that users have to fix are detected during term validation, whereas the other $3$ components succeed most of the time.
By isolating term validation, we can provide substantial UI features, which are especially helpful when term validation does not succeed.
Moreover, many UI features do not depend on term validation such as auto-completion or search.

Thirdly, it opens up an alternative way to connect an existing kernel with our UI in those cases where it is not possible to have the kernel expose the AST.
Then it can be feasible to reimplement the other $3$ components from scratch, but it will usually be impossible to reimplement term validation.

We explain the $2\times 2$ components in the following, using our kernel for LF as a running example.
To simplify the description below, we point out two \textbf{general aspects} globally that apply to all $2\times 2$ components.
Firstly, to maximize \textbf{extensibility}, both individual kernels as well as extensions of our example kernel are supplied to \jmmt as plugins.
Thus, kernels can be developed and used without rebuilding \jmmt itself.

Secondly, all $2\times 2$ components can \textbf{report errors} to \jmmt and can always recover from errors by returning partial or no results.
For example, term parsing may return a term with some unparsed sub-terms, for which errors are reported.

\subsection{Structure Parsing}\label{sec:sp}

Structure parsing takes a source document and returns an abstract syntax tree (AST) in the \mmt language.
Alternatively, the AST can be returned as XML (using the \omdoc format \cite{omdoc}), which is useful if kernels are implemented in other programming languages.
The terms occurring in the source can remain unparsed: \jmmt produces a parsing unit for each unparsed object, which is passed on to the term parsing component.

\begin{example}[A Keyword-Based Parser]
We implement a straightforward structure parser based on keywords.
First it splits a file into declarations based on a few reserved characters and computes their source references.
Then the parsing of each declaration is relegated to an appropriate plugin, which is selected based on the keyword.

We choose the ASCII characters 28-31 as separators (the green boxes in Fig.~\ref{fig:jedit}).
For entering them, we provide \jedit actions that users can bind to, e.g., any key combination.
Thus, existing term parsers can be reused easily no matter what special characters are used in the terms.

Our structure parser works at the \mmt level, and we do not have to customize it for LF at all.
\end{example}

\subsection{Structure Validation}\label{sec:sv}

Structure validation is a fixed algorithm, that implements the semantics of \mmt, in particular the module system \cite{RK:mmt:10}.
Thus, it is not necessary to change the implementation in specific kernels.

Because \mmt is parametric in the underlying type system, the terms cannot be validated generically.
Instead, structure validation produces one validation unit for each term and passes them on to the term validator.
In particular, the constant declaration $c:A=t$ in a theory $T$ gives rise to two validation units:
$\ctp{c}$ validates the judgment $\isuniv{T}{A}$, and $\cdef{c}$ validates the judgment $\oftype{T}{t}{A}$.

Even though this algorithm is fixed, it is still extensible because it implements the structure level extension mechanism we presented in \cite{HKR:extending:12}.
This is general enough to express, e.g., all toplevel declarations of Mizar (as we show in \cite{IKRU:mizar:11}) or the type definition principle of HOL systems.

\subsection{Term Parsing}\label{sec:tp}

A \textbf{parsing unit} is a tuple of
\begin{inparaenum}
	\item a string $S$ that is to be parsed into a term,
	\item the source reference $R$ of $S$ (if returned by the structure parser),
	\item the theory $T$ in which $S$ occurs.
\end{inparaenum}
Term parsing takes as input a parsing unit and returns an \mmt term $t$ in context $\Gamma$.
The role of $\Gamma$ is to declare meta-variables for unknown sub-terms that are to be found during validation.
For example, the term parser may generate a meta-variable for the omitted type of a bound variable.
More generally, meta-variables can be used to represent proof that still have be found and inserted.

In \mmt, each sub-term may carry its own source reference.
Thus, the term parsing algorithm can return fine-granular source references for all sub-terms.
The \mmt implementation makes sure that source references are ignored when programs inspect terms (e.g., to check identity or for pattern-matching), but are carried along when programs transform terms by substitution or rewriting.

\begin{example}[A Notation-Based Parser]
\jmmt includes a term parser, which uses the \mmt notations of all constants that are imported into $T$.
It returns source references for every sub-term and inserts fresh meta-variables for omitted variable types and implicit arguments.
(An argument position is considered implicit if it is not mentioned in the notation.)

In fact, our term parser is a bit more general: It also supports notations with precedences, argument sequences, and bound variable sequences.

Our term parser works at the \mmt level, and we do not have to customize it for LF -- we only have to import the theory from Ex.~\ref{ex:lf}.
\end{example}

\subsection{Term Validation}\label{sec:tv}

A \textbf{validation unit} consists of
\begin{inparaenum}
  \item a context $\Gamma$ (as returned by the term parser),
  \item a judgment (as in Fig.~\ref{fig:judge}) whose terms may use the free variables of $\Gamma$.
\end{inparaenum}

A term validator takes a validation unit; it verifies the judgment and returns a substitution for the variables in $\Gamma$, i.e., it finds the unknown parts of the term.
This has the advantage that the validated term has the same structure as the parsed term so that the UI can present both at once.
Alternatively, it may be reasonable that validation returns a whole new term, e.g., its normal form.

It is important that we validate at a per-term basis and not at a per-declaration basis.
As described above, our structure validator produces two independent validation units $\ctp{c}$ and $\cdef{c}$ for a declaration $c:A=t$.
For most type theories, $\cdef{c}$ implies $\ctp{c}$ so that validating both seems redundant.
But this design is crucial for change management, as we have seen in Sect.~\ref{sec:moc}.

\begin{example}[A Rule-Based Validator]\label{ex:tv}
We have developed a term validator for \mmt terms.
It reconstructs the unknown variables from $\Gamma$ by using an algorithm similar to the one of Twelf \cite{twelf}.
Moreover, type checking can be performed modulo rewriting according to user-declared equalities (which are assumed to be confluent).
This term validator is described in detail in \cite{rabe:mmttypetheory:14}.

In order to maximize reuse, we only implement certain structural rules like congruence and lookup.
We use the heads of complex terms to relegate to language-specific rules, which can be provided by other plugins.
\end{example}

\begin{example}[A Term Validator for LF]\label{ex:tvlf}
	We instantiate the term validator from Ex.~\ref{ex:tv} with LF by providing LF-specific rules.
	These are inference rules that infer the types of terms with head $\type$, $\lflambda$, $\lfPi$, or $\lfapply$; a checking rule for judgments of the form $\oftype{T}{f}{\{x:A\}B}$; an equality rule for judgments of the form $\isequal[\{x:A\}B]{T}{t}{t'}$; a rewrite rule to turn $A\to B$ into $\{\_:A\}B$ and a rewrite rule for $\beta$-reducible terms of the form $([x:A]t)a$.
	
	Moreover, we implement a solution rule, which is used to determine the values of the meta-variables in $\Gamma$: It applies to judgments of the form $\isequal[A]{T}{X\;x_1\;\ldots\;n_n}{E}$, where $X$ is a meta-variable from $\Gamma$, and the $x_i$ are distinct bound variables, and solves $X$ as $[x_1,\ldots,x_n]E$.
\end{example}

Notably, in terms of lines of code, the LF-specific rules from Ex.~\ref{ex:tvlf}, the generic term validator from Ex.~\ref{ex:tv}, our plugin for \jedit together with \mmt, and the whole \jedit code are roughly related like $1:5:500:5000$.
This indicates the potential synergies of logic-independent IDEs.

\section{Conclusion}\label{sec:conc}
  We have developed \jmmt, a generic user interface (UI) for formal logics, by combining the \mmt representation language and the \jedit text editor.
Our UI differs from most current ones by employing a high-level abstraction layer that permits a high-level UI.
This permits a number of advanced features including auto-completion, error highlighting, search, navigation, and change management.
Future work can benefit from the existing abstraction layer and extend these features substantially.

But we have to pay for this feature richness: Our IDE cannot be directly used with existing logic implementations.
Even though these largely include the algorithms that our abstraction layer assumes, these algorithms are not easy to expose.
Depending on the individual system, this may require even partial redesigns.

Among IDEs for \emph{existing} logic implementations, our closest relative is Isabelle/\jedit \cite{isabelle_jedit}, which was developed independently of ours.
Both systems share the idea of a high-level UI based on \jedit as well as the basic features (Sect.~\ref{sec:basic}).
The main difference is that \jmmt is a \jedit plugin, whereas Isabelle/\jedit forks a few parts of the \jedit source and is integrated with the Isabelle system.
Consequently, Isabelle/\jedit can offer additional Isabelle-specific features, in particular in regard to theorem proving.
\jmmt, on the other hand, offers several advanced features and focuses on logic-independence and extensibility.

\emph{New} logic implementations can easily reuse \jmmt by providing plugins that instantiate our abstraction layer.
We demonstrated that by developing a mature implementation for the logical framework LF from scratch.
Moreover, new logic implementations can customize our LF implementation by adding operators, notations, or typing rules -- in that case, developers obtain the double benefit of a strong UI at very small implementation cost.
Additionally, they can benefit from the existing \mmt infrastructure, in particular the module system.


\bibliographystyle{alpha}

\end{document}